\documentclass[sigconf]{acmart}
\usepackage{lipsum}

\AtBeginDocument{%
  \providecommand\BibTeX{{%
    \normalfont B\kern-0.5em{\scshape i\kern-0.25em b}\kern-0.8em\TeX}}}

\copyrightyear{2023} 
\acmYear{2023} 
\setcopyright{acmlicensed}\acmConference[WWW '23 Companion]{Companion Proceedings of the ACM Web Conference 2023}{April 30-May 4, 2023}{Austin, TX, USA}
\acmBooktitle{Companion Proceedings of the ACM Web Conference 2023 (WWW '23 Companion), April 30-May 4, 2023, Austin, TX, USA}
\acmPrice{15.00}
\acmDOI{10.1145/3543873.3587373}
\acmISBN{978-1-4503-9419-2/23/04}

\usepackage{subcaption}
\usepackage{mwe}

\begin{document}
\newcommand{\Secref}[1]{Section~\ref{#1}}
\newcommand{\Figref}[1]{Fig.~\ref{#1}}

\title{Measuring and Detecting Virality on Social Media: The Case of Twitter's Viral Tweets Topic}


\author{Tuğrulcan Elmas}
\affiliation{
  \institution{Indiana University}
  \city{Bloomington}
  \country{U.S.A.}}
\email{telmas@iu.edu}

\author{Stephane Selim}
\affiliation{
  \institution{EPFL}
  \city{Lausanne}
  \country{Switzerland}}
\email{stephan.selim@epfl.ch}

\author{Célia Houssiaux}
\affiliation{
  \institution{EPFL}
  \city{Lausanne}
  \country{Switzerland}}
\email{celia.houssiaux@epfl.ch}






\begin{abstract}

Social media posts may go viral and reach large numbers of people within a short period of time. Such posts may threaten the public dialogue if they contain misleading content, making their early detection highly crucial. Previous works proposed their own metrics to annotate if a tweet is viral or not in order to automatically detect them later. However, such metrics may not accurately represent viral tweets or may introduce too many false positives. In this work, we use the ground truth data provided by Twitter's "Viral Tweets" topic to review the current metrics and also propose our own metric. We find that a tweet is more likely to be classified as viral by Twitter if the ratio of retweets to its author's followers exceeds some threshold. We found this threshold to be 2.16 in our experiments. This rule results in less false positives although it favors smaller accounts. We also propose a transformers-based model to early detect viral tweets which reports an F1 score of 0.79. The code and the tweet ids are publicly available at: https://github.com/tugrulz/ViralTweets

\end{abstract}

\maketitle

\section{Introduction}

\begin{figure}[!htb]
    \centering
    \includegraphics[width = \columnwidth]{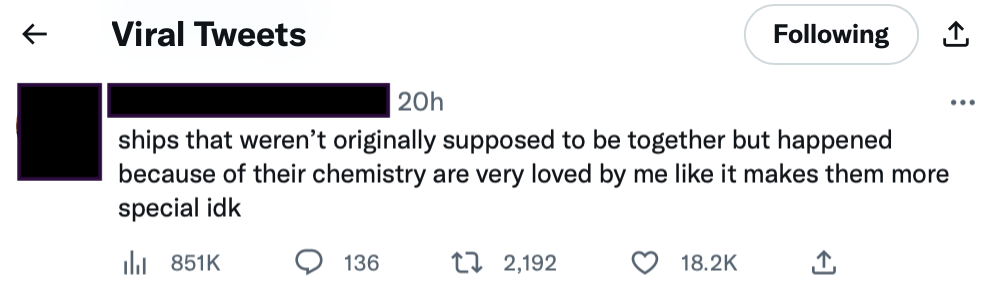}
    \caption{An example of a tweet listed under Viral Tweets topic page. The tweet was viewed by 851,000 users in 20 hours despite that the account had only 1000 followers.}
    \label{fig:exampleviral}
\end{figure}

Social media platforms have the power to shape public opinion and spark widespread conversations in a matter of seconds. On those platforms, the posts may go viral, reaching thousands, if not millions, of people within a short period of time even though their authors were not popular initially. Those who seek to maximize influence may craft their campaigns so that they will go viral through such posts. Unfortunately, adversaries may also adopt such an approach and craft viral and misleading content such as fake news and conspiracy theories. Understanding viral posts may help early detection of such content. This is especially crucial for fact-checking: once a claim goes viral, the impact of fact-checking diminishes. On the other hand, it is infeasible for fact-checkers to proactively fact-check every claim. Early detection of viral posts may help them prioritize claims to fact-check~\cite{guo2022survey}. Moreover, early detection may facilitate analyzing viral posts before they are removed by the adversaries or the platforms~\cite{elmas2023impact}.

Past work on predicting viral social media posts mainly focused on Twitter. To build a ground truth set of viral tweets, they used human labeling~\cite{maldonado2021virality} or define their own measures based on public metrics as proxies for virality~\cite{rameez2022viralbert, Jenders-2013}. For instance, Jenders et al.~\cite{Jenders-2013} consider tweets with a number of retweets more than a threshold as "viral". However, such proxies may be too restrictive (i.e. low recall on viral posts) or too lenient (many non-viral posts are labeled as viral) Furthermore, they fell short of accounting for the tweets' impact on a network, i.e., a tweet that is retweeted a lot may not always reach more users. Furthermore, public metrics may be manipulated (e.g., by bots~\cite{elmas2021ephemeral,elmas2022characterizing}) and may result in misclassifying non-viral tweets. A more convincing approach is to predict the viral tweets \textit{of a given user}, which will control the effect of users' network and likeliness to use bots. 

In 2021, Twitter launched the "Viral Tweets" topic which discloses the tweets that went viral on the platform (e.g., \Figref{fig:exampleviral}) and provides reliable ground truth data. In this study, we survey the existing measures and evaluate their ability to capture the viral tweets provided by Twitter. We also propose our own metric based on the retweet and followers of the users and show that it is more precise than the previous methods even though it requires fewer data. We propose a transformers-based method to detect Twitter's viral tweets without relying on the tweets' or users' public metrics to facilitate early detection of viral tweets. Our work will facilitate future research on measuring and detecting virality on social media.

\section{Related Works}

Previous works define and predict viral tweets based on their definitions. Jenders et al.~\cite{Jenders-2013} employ tweet metadata features to predict the likelihood of a tweet becoming "viral". Viral tweets are defined as tweets that are retweeted at least T times, with T chosen as 50, 100, 500, or 1000. Maldonado-Sifuentes et al.~\cite{maldonado2021virality} employed RoBERTa for predicting the virality of a tweet using a corpus of 5000 tweets annotated by humans. Zadeh et al.~\cite{zadeh2022can} propose and test a framework based on multivariate Hawkes processes for predicting the popularity (defined as the sum of retweets, replies, and likes) of Twitter posts by brands, using regression rather than classification. Garimella et al.~\cite{garimella2019hot} defined the hot streaks (a series of viral tweets in a period) as the tweets which have more retweets than the 90\% of the other tweets of the user.

Our work differs from these in two aspects. First, rather than defining viral tweets by ourselves based on proxies which may not be reliable, we tackle the prediction of viral tweets disclosed by Twitter itself. These viral tweets did or have the potential to spread to a wide range of Twitter users. This is because we assume that the platform is able to better define and model viral social media posts due to the internal metrics it has. Secondly, we create and evaluate by a setting that helps check-worthiness estimation in the fact-checking pipeline and aid fact-checkers, which controls the users while predicting the viral tweets.

Other works on viral tweets involve characterizing them. Samuel et al.~\cite{samuel2020message} analyzed a dataset of over one million Tweets to understand the key drivers of successful information exchange and message diffusion on Twitter, focusing on endogenous and exogenous dimensions and providing insights and an early-stage model for explaining tweet performance. Sprejer et al.~\cite{sprejer2021influencer} investigates the virality of radical right content from 35 radical right influencers. They find that both influencer and content-level factors, including the number of followers, type of content, length, and toxicity of the content, and requests for retweets, are important for engagement with tweets. Hoang et al.~\cite{hoang2011modeling} study the virality of socio-political tweet content in Singapore's 2011 general election by collecting tweet data from 20,000 Singaporean users and introducing several quantitative indices to measure the virality of tweets that are retweeted. They identify the most viral messages and the users behind them in the election and explain their behavior. Hasan et al.~\cite{hasan2022impact} investigates the effects of virality on users' subsequent behaviors and long-term visibility on the platform using a dataset of tweeting activities and follower graph changes for 17,157 scientists on Twitter. Gurjar et al~\cite{gurjar2022effect} propose a framework to examine changes in user activity and the survival duration of effects associated with popularity shocks. Elmas et al.~\cite{elmas2020misleading} show that viral social media accounts may be sold and repurposed for malicious purposes later.

\section{Data}

\begin{figure*}[!htb]
\subfloat{\includegraphics[width=0.44\linewidth]{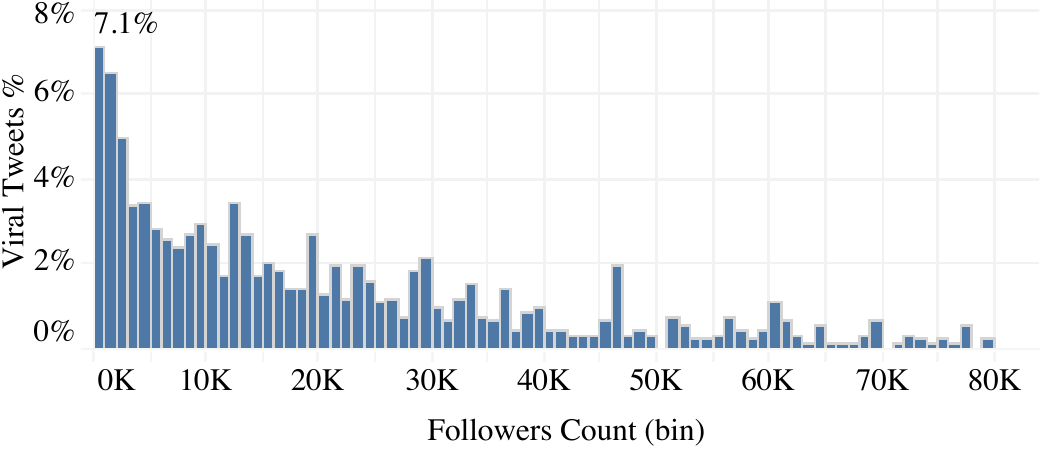}
\label{mention:fol}}
\hfill
\subfloat{\includegraphics[width=0.44\linewidth]{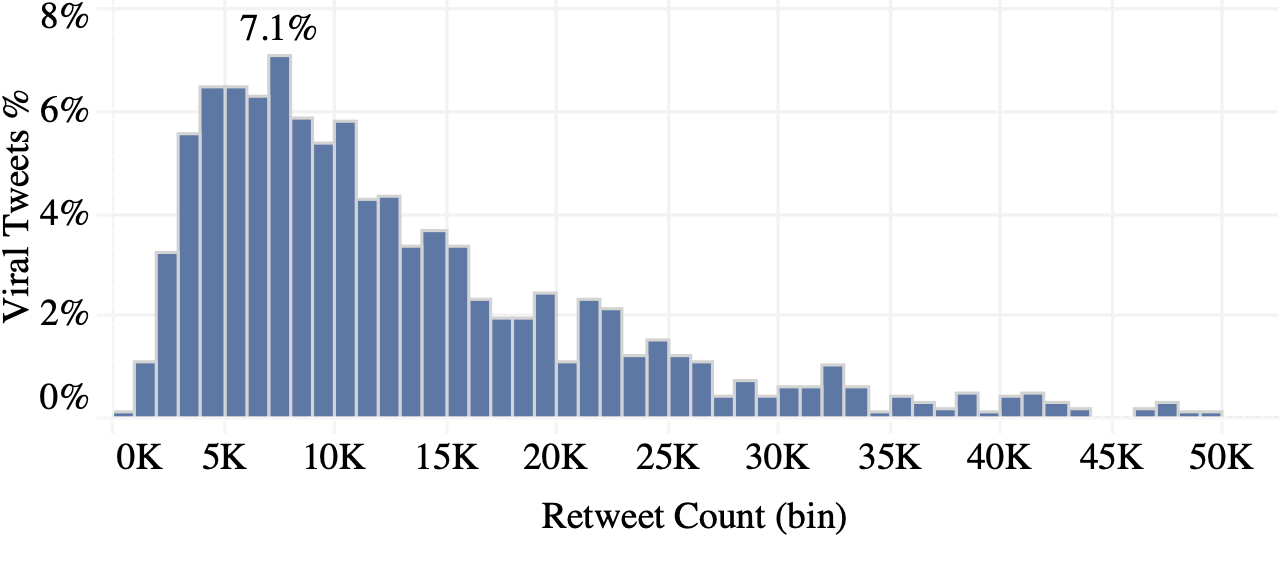}
\label{mention:rt}}
\caption{The retweet counts of the tweets and the follower counts of the authors in bins with size 1000. We observe that the data is skewed left, towards unpopular users with retweet counts less than 10,000.}
\label{fig:histograms}
\end{figure*}

\textit{Topics} is a Twitter feature where users can subscribe to a feed of interest such as sports teams, art, food, etc. Twitter curates viral tweets (defined as tweets that are "Popular now") under the "Viral Tweets" topic. Unfortunately, the Twitter API is yet to provide an API endpoint to collect tweets under a topic. Thus, we scraped 1008 tweets from this page between October 2022 and November 2022 and collected their ids. We identified 814 users who authored viral tweets. 89 users posted more than one viral tweet. We collected the last 3200 tweets (which is the limit enforced by the API) of those users to build the dataset of non-viral tweets. We excluded retweets. We collected additional 1,137,050 tweets through this process.

To explore the datasets with respect to public metrics to measure virality, we show the distribution of the retweet counts of the viral tweets and the follower counts of their authors in \Figref{fig:histograms}. We observe that the viral tweets are sourced mostly from unpopular profiles and retweet counts are usually below 10,000. 

\section{Measuring Virality}

Viral tweets are tweets that spread on Twitter and reach a large number of users in a short period. Since external researchers cannot reliably model how many users a tweet reaches, they generally measure virality using proxies, mainly based on the number of retweets. We now survey these measures.
\\\noindent\textbf{RT > T:} The number of retweets should be greater than a predefined hard threshold. The pitfall of this measure is that many users with a high number of followers may acquire retweets higher than such a threshold from their own network, even though they did not go viral. The measure relies only on the retweet count of the tweet as the data. Thus, we propose the following measure that takes the overall performance of the users' tweets:

\noindent\textbf{RT / Med. RT or RT / Avg. RT:} The number of retweets divided by the median or the average retweet count of the users' tweets should be greater than some threshold. In the former case, tweets with zero retweets are not taken into account. The measure requires the users' timeline, i.e., recent tweets to compute the median or the average. A similar approach was proposed by Garimella et al.~\cite{garimella2019hot}:

\noindent\textbf{RT Percentile:} The number of retweets should be greater than the kth percentile of the tweets' retweet counts. The metric assumes that the user will have a fixed number of viral tweets, which may be problematic if the user does not have any viral tweets, or has many viral tweets. This metric also requires users' timelines.

\noindent\textbf{RT / Followers:} Instead of profiling the users using their timeline, we instead use their number of followers. Thus, we normalize the number of retweets the tweet has by dividing the number of followers the user has. 

\noindent\textbf{log(RT / Followers):} The number of retweets may not increase linearly with the number of followers. To account for this, we compute the natural logarithm of the previous value. We discarded similar measures such as log(RT)/Followers and RT/log(Followers) as they were outperformed by this measure in the experiments.

\noindent\textbf{Influence Score:} The metric by Maldonado-Sifuentes et al.~\cite{maldonado2021virality}: 

\begin{equation}
    inf_{twt} = \dfrac{gd*(Ar+f)}{wr*(Ad+h)}
\end{equation}

where $w$ is followers count, $d$ is the followings count, $r$ is the retweets count, $f$ is the number of favorites. $A$ is a constant, which was set to 10 by the authors. $g = r + f$ and $h = w - d$. 

\begin{figure*}
    \centering
    \includegraphics[width=0.875\textwidth]{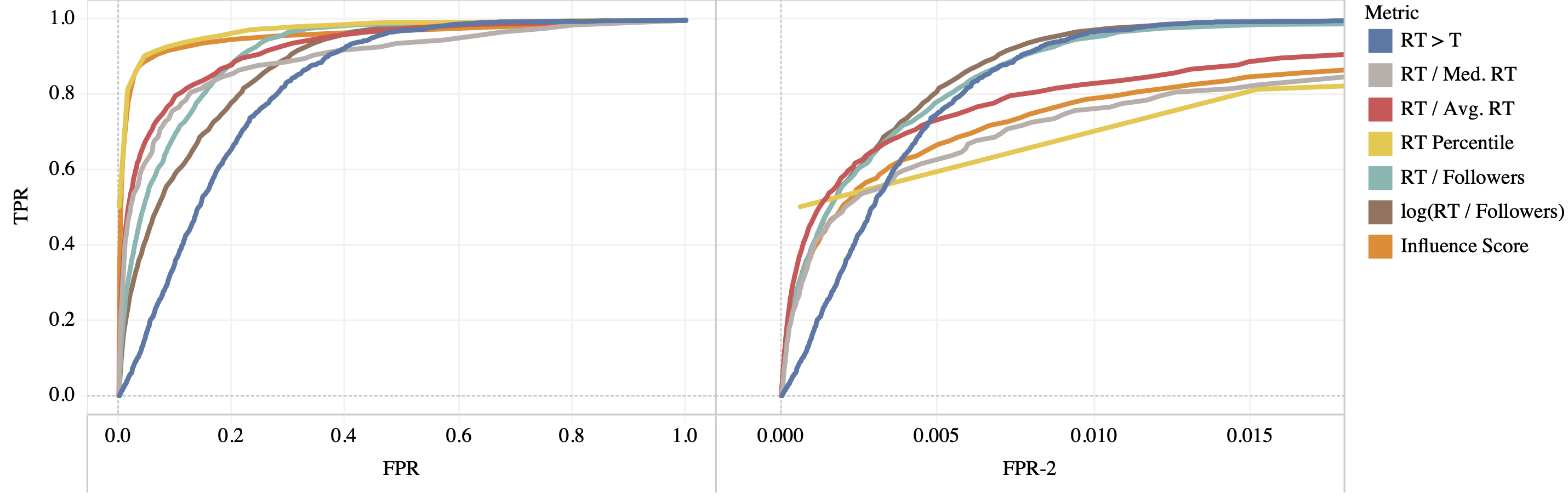}
    \caption{The ROC curves of different metrics, drawn with the two different approaches of computing FPRs. }
    \label{fig:metrics}
\end{figure*}

We run these metrics on all the tweets in our dataset to automatically classify if a tweet is viral or not. We define true positives as tweets that are classified as viral by the metric and amplified as viral tweets by Twitter in the Viral Tweets topic. We adopt two approaches to define the false positives set. The first is the smallest set of tweets that the given metric classifies as viral when the metric reaches a 100\% true positive rate. In this case, each metric will have a different false negative set. The second is the largest of tweets, which will be all 1,3m non-viral tweets in the dataset. We experiment with different thresholds and percentiles to compute true positive rates and false positive rates that range between 0 and 1.0 by 0.01 steps. \Figref{fig:metrics} show the ROC curves for all the metrics.

We compute the AUC with the former approach to compute FPR using the standard method as the FPR is already scaled between 0 and 1. For the latter approach, as the FPRs are very low, the resulting AUCs are very high and close to each other. Thus, we constrain the FPR to be between 0 and 0.016 as we observe that the ROC curves are mostly stable by this FPR rate. We then rescale the FPRs to 0 and 1 and compute the AUC, which we name AUC-2. We also compute the number of false positives at TPR = 0.95 for each metric to show how lenient they are, i.e., how many new viral tweets they introduce. Table \ref{tab:auc} shows the results. We observe that the influence score does better in AUC, but performs poorly in AUC-2. This is because it is very lenient and introduces too many false positives, 225k when TPR = 0.95. Although a hard threshold (RT > T) is not as lenient, it results in high FPR to achieve a high TPR. Meanwhile, both RT / Followers and log(RT / Followers) achieve well on both metrics, but the latter is less lenient and has a higher harmonic mean of AUC and AUC-2. To put these into practice, the best threshold for log(RT/Followers) is 0.772, i.e., 2160 retweets per 1000 followers. The best hard threshold ($RT > T$) is 3088 retweets. The former rule favors smaller accounts while the latter favors popular ones.




\begin{table}[]
\caption{Evaluation results for each virality metric}
\begin{tabular}{|l|l|l|l|l|}
\hline
Metric                  & Data Required & AUC & AUC-2 & \#Viral \\ \hline
RT >    T~\cite{Jenders-2013,zadeh2022can}       & Tweet Only          &  0.82     &    0.78   & 12,439                       \\ \hline
RT /  Med. RT & Timeline &    0.90   &  0.67     & 85,185                       \\ \hline
RT /  Avg. RT & Timeline &  0.93     &   0.75    & 39,923                       \\ \hline
RT Percentile~\cite{garimella2019hot}           & Timeline &   0.84    &   0.64    & 203,539                      \\ \hline
RT / Followers          & Profile  &   0.92    &   0.82    & 14,007                       \\ \hline
log(RT / Followers)     & Profile  &    0.88   & \textbf{0.86}     & 12,034                       \\ \hline
Influence Score~\cite{maldonado2021virality}         & Profile &   \textbf{0.96}    &   0.70    & 225,314                      \\ \hline
\end{tabular}
\label{tab:auc}
\end{table}

\section{Detection}

\subsection{Motivation \& Problem}

Our goal in predicting viral tweets is to aid fact-checking by proactively detecting tweets that may spread to many social media users in a short period and make an impact. Thus, we do not use any engagement information (e.g., like count) and focus on content features only. We assume that fact-checkers track a set of users that may share misleading content in real-time, detect if they have a claim in their tweets, and estimate the check worthiness of those claims. Since there may be too many tweets to detect check worthiness or too many check-worthy tweets to fact-check, it may be a better strategy to track users that may go viral and prioritize their tweets. Thus, we formulate our problem as follows: "Given the tweets from a set of users, which tweets are likely to go viral?". 

We set the data according to this problem and constrained it to the viral tweets and their authors' non-viral tweets on the same day. We only used the tweets in English. This leaves us with 787 viral tweets and 15,904 non-viral tweets. We randomly sample from the non-viral dataset to achieve a balanced dataset. We use a balanced training set of 1,260 tweets and a test set of 314 tweets. 

\subsection{Feature Engineering}

We mainly employ text content as the feature and use transformer-based language models to represent it. We also use additional features that the language models may not model. In our experiment, we observe that such additional features slightly increase our model's performance. Those features are the boolean features that whether the tweet contains media, hashtags, mentions, has positive sentiment, negative sentiment, and if it is sourced from a verified account. We used \textit{distilbert-base-uncased-finetuned-sst-2-english} model to compute sentiment, which yields if the tweet is positive and negative with a confidence score. We assign the sentiment of the tweet if the confidence score was higher than 0.7. Table \ref{tab:top_features} summarizes the features. 

\begin{table}[]
  \caption{The features, their share (or mean) among viral and non-viral tweets. The differences are statistically significant (p < 0.05) except for the "Contains Mentions" feature. }
  \begin{tabular}{lllll}
    \hline
    Feature & Viral & Non-Viral & Diff \\\hline
    Contains Media & \%62.1 & \%21.7 & \%40.4 \\ 
    Contains Hashtags & \%5.85 & \%3.03 & \%2.82 \\
    From Verified Account & \%5.46 & \%7.24 & \%1.78 \\
    Positive Sentiment & \%25.2 & \%40 & \%14.8  \\
    Negative Sentiment & \%74.8 & \%60 & \%14.8 \\
    Contains Mentions & \%42.76 & \%41.12 & \%1.64\\
    Mean Tweet Length & 88.3 & 64.9 & 23.4  \\

    \hline
  \end{tabular}

  \label{tab:top_features}
\end{table}

\subsection{Experimental Results}

We used the following transformers-based language models using HuggingFace: BERT-Base~\cite{devlin2018bert}, RoBERTa~\cite{liu2019roberta}, TinyBERT~\cite{jiao2019tinybert}, and BERTweet~\cite{nguyen2020bertweet}. We only used the case-sensitive models as we observe that users use upper case when they want to put emphasis on a certain part of their tweets. We experimented with models that rely only on the text content and models that concatenate the features we created to text features. We evaluated the models using Accuracy, Precision, Recall, and F1. Table \ref{tab:results} shows the result. We observe that BERTweet yields the best F1 and using the extra features increases it by 0.027.

\begin{table}[htbp]
  \centering
  \caption{Results using the plain model and the model with extra features marked with*}
  \begin{tabular}[c]{|l||l|l|l|l|l|l|}
    \hline
    Model & Prec & Prec* & Recall & Recall* & F1 & F1* \\
    \hline
    BERT-Base & 0.670 & 0.666 & 0.764 & 0.815 & 0.714 & 0.734 \\
    RoBERTa & 0.704 & 0.681 & 0.834 & 0.860 & 0.764 & 0.761 \\
    TinyBERT & 0.690 & 0.668 & 0.834 & 0.885 & 0.755 & 0.762 \\
    BERTweet & 0.717 & 0.740 & 0.822 & 0.854 & 0.766 & \textbf{0.793} \\
    \hline
  \end{tabular}

  \label{tab:results}
\end{table}
\section{Conclusion and Future Work}

This study improves virality understanding on social media by testing existing metrics on reliable data, proposing a new metric, and predicting viral tweets. Our analysis predates Twitter's view count disclosure policy. However, our virality metrics may benefit the works on other social media platforms which do not disclose such data. Additionally, our work on predicting viral tweets using language models may inspire future work which automatically generates content that is likely to go viral, which may help experts and fact-checkers to create content that better resonates with the public. Furthermore, viral tweets tend to have more media content, and further research could focus on modeling this.

\noindent\textbf{Ethical Disclosure:} We only used the data from public profiles amplified by Twitter. We only disclose the tweet ids from the data.





\bibliographystyle{ACM-Reference-Format}
\bibliography{sample-base}


\begin{thebibliography}{19}


\ifx \showCODEN    \undefined \def \showCODEN     #1{\unskip}     \fi
\ifx \showDOI      \undefined \def \showDOI       #1{#1}\fi
\ifx \showISBNx    \undefined \def \showISBNx     #1{\unskip}     \fi
\ifx \showISBNxiii \undefined \def \showISBNxiii  #1{\unskip}     \fi
\ifx \showISSN     \undefined \def \showISSN      #1{\unskip}     \fi
\ifx \showLCCN     \undefined \def \showLCCN      #1{\unskip}     \fi
\ifx \shownote     \undefined \def \shownote      #1{#1}          \fi
\ifx \showarticletitle \undefined \def \showarticletitle #1{#1}   \fi
\ifx \showURL      \undefined \def \showURL       {\relax}        \fi
\providecommand\bibfield[2]{#2}
\providecommand\bibinfo[2]{#2}
\providecommand\natexlab[1]{#1}
\providecommand\showeprint[2][]{arXiv:#2}

\bibitem[Devlin et~al\mbox{.}(2018)]%
        {devlin2018bert}
\bibfield{author}{\bibinfo{person}{Jacob Devlin}, \bibinfo{person}{Ming-Wei
  Chang}, \bibinfo{person}{Kenton Lee}, {and} \bibinfo{person}{Kristina
  Toutanova}.} \bibinfo{year}{2018}\natexlab{}.
\newblock \showarticletitle{Bert: Pre-training of deep bidirectional
  transformers for language understanding}.
\newblock \bibinfo{journal}{\emph{arXiv preprint arXiv:1810.04805}}
  (\bibinfo{year}{2018}).
\newblock


\bibitem[Elmas(2023)]%
        {elmas2023impact}
\bibfield{author}{\bibinfo{person}{Tu{\u{g}}rulcan Elmas}.}
  \bibinfo{year}{2023}\natexlab{}.
\newblock \showarticletitle{The Impact of Data Persistence Bias on Social Media
  Studies}.
\newblock \bibinfo{journal}{\emph{arXiv preprint arXiv:2303.00902}}
  (\bibinfo{year}{2023}).
\newblock


\bibitem[Elmas et~al\mbox{.}(2022)]%
        {elmas2022characterizing}
\bibfield{author}{\bibinfo{person}{Tu{\u{g}}rulcan Elmas},
  \bibinfo{person}{Rebekah Overdorf}, {and} \bibinfo{person}{Karl Aberer}.}
  \bibinfo{year}{2022}\natexlab{}.
\newblock \showarticletitle{Characterizing Retweet Bots: The Case of Black
  Market Accounts}. In \bibinfo{booktitle}{\emph{Proceedings of the
  International AAAI Conference on Web and Social Media}},
  Vol.~\bibinfo{volume}{16}. \bibinfo{pages}{171--182}.
\newblock


\bibitem[Elmas et~al\mbox{.}(2020)]%
        {elmas2020misleading}
\bibfield{author}{\bibinfo{person}{Tu{\u{g}}rulcan Elmas},
  \bibinfo{person}{Rebekah Overdorf}, \bibinfo{person}{{\"O}mer~Faruk
  Akg{\"u}l}, {and} \bibinfo{person}{Karl Aberer}.}
  \bibinfo{year}{2020}\natexlab{}.
\newblock \showarticletitle{Misleading repurposing on twitter}.
\newblock \bibinfo{journal}{\emph{arXiv preprint arXiv:2010.10600}}
  (\bibinfo{year}{2020}).
\newblock


\bibitem[Elmas et~al\mbox{.}(2021)]%
        {elmas2021ephemeral}
\bibfield{author}{\bibinfo{person}{Tu{\u{g}}rulcan Elmas},
  \bibinfo{person}{Rebekah Overdorf}, \bibinfo{person}{Ahmed~Furkan
  {\"O}zkalay}, {and} \bibinfo{person}{Karl Aberer}.}
  \bibinfo{year}{2021}\natexlab{}.
\newblock \showarticletitle{Ephemeral astroturfing attacks: The case of fake
  twitter trends}. In \bibinfo{booktitle}{\emph{2021 IEEE European Symposium on
  Security and Privacy (EuroS\&P)}}. IEEE, \bibinfo{pages}{403--422}.
\newblock


\bibitem[Garimella and West(2019)]%
        {garimella2019hot}
\bibfield{author}{\bibinfo{person}{Kiran Garimella} {and}
  \bibinfo{person}{Robert West}.} \bibinfo{year}{2019}\natexlab{}.
\newblock \showarticletitle{Hot streaks on social media}. In
  \bibinfo{booktitle}{\emph{Proceedings of the international AAAI conference on
  web and social media}}, Vol.~\bibinfo{volume}{13}. \bibinfo{pages}{170--180}.
\newblock


\bibitem[Guo et~al\mbox{.}(2022)]%
        {guo2022survey}
\bibfield{author}{\bibinfo{person}{Zhijiang Guo}, \bibinfo{person}{Michael
  Schlichtkrull}, {and} \bibinfo{person}{Andreas Vlachos}.}
  \bibinfo{year}{2022}\natexlab{}.
\newblock \showarticletitle{A survey on automated fact-checking}.
\newblock \bibinfo{journal}{\emph{Transactions of the Association for
  Computational Linguistics}}  \bibinfo{volume}{10} (\bibinfo{year}{2022}),
  \bibinfo{pages}{178--206}.
\newblock


\bibitem[Gurjar et~al\mbox{.}(2022)]%
        {gurjar2022effect}
\bibfield{author}{\bibinfo{person}{Omkar Gurjar}, \bibinfo{person}{Tanmay
  Bansal}, \bibinfo{person}{Hitkul Jangra}, \bibinfo{person}{Hemank Lamba},
  {and} \bibinfo{person}{Ponnurangam Kumaraguru}.}
  \bibinfo{year}{2022}\natexlab{}.
\newblock \showarticletitle{Effect of Popularity Shocks on User Behaviour}. In
  \bibinfo{booktitle}{\emph{Proceedings of the International AAAI Conference on
  Web and Social Media}}, Vol.~\bibinfo{volume}{16}. \bibinfo{pages}{253--263}.
\newblock


\bibitem[Hasan et~al\mbox{.}(2022)]%
        {hasan2022impact}
\bibfield{author}{\bibinfo{person}{Rakibul Hasan}, \bibinfo{person}{Cristobal
  Cheyre}, \bibinfo{person}{Yong-Yeol Ahn}, \bibinfo{person}{Roberto Hoyle},
  {and} \bibinfo{person}{Apu Kapadia}.} \bibinfo{year}{2022}\natexlab{}.
\newblock \showarticletitle{The Impact of Viral Posts on Visibility and
  Behavior of Professionals: A Longitudinal Study of Scientists on Twitter}. In
  \bibinfo{booktitle}{\emph{Proceedings of the International AAAI Conference on
  Web and Social Media}}, Vol.~\bibinfo{volume}{16}. \bibinfo{pages}{323--334}.
\newblock


\bibitem[Hoang et~al\mbox{.}(2011)]%
        {hoang2011modeling}
\bibfield{author}{\bibinfo{person}{Tuan-Anh Hoang}, \bibinfo{person}{Ee-Peng
  Lim}, \bibinfo{person}{Palakorn Achananuparp}, \bibinfo{person}{Jing Jiang},
  {and} \bibinfo{person}{Feida Zhu}.} \bibinfo{year}{2011}\natexlab{}.
\newblock \showarticletitle{On modeling virality of twitter content}. In
  \bibinfo{booktitle}{\emph{International conference on Asian digital
  libraries}}. Springer, \bibinfo{pages}{212--221}.
\newblock


\bibitem[Jiao et~al\mbox{.}(2019)]%
        {jiao2019tinybert}
\bibfield{author}{\bibinfo{person}{Xiaoqi Jiao}, \bibinfo{person}{Yichun Yin},
  \bibinfo{person}{Lifeng Shang}, \bibinfo{person}{Xin Jiang},
  \bibinfo{person}{Xiao Chen}, \bibinfo{person}{Linlin Li},
  \bibinfo{person}{Fang Wang}, {and} \bibinfo{person}{Qun Liu}.}
  \bibinfo{year}{2019}\natexlab{}.
\newblock \showarticletitle{Tinybert: Distilling bert for natural language
  understanding}.
\newblock \bibinfo{journal}{\emph{arXiv preprint arXiv:1909.10351}}
  (\bibinfo{year}{2019}).
\newblock


\bibitem[Liu et~al\mbox{.}(2019)]%
        {liu2019roberta}
\bibfield{author}{\bibinfo{person}{Yinhan Liu}, \bibinfo{person}{Myle Ott},
  \bibinfo{person}{Naman Goyal}, \bibinfo{person}{Jingfei Du},
  \bibinfo{person}{Mandar Joshi}, \bibinfo{person}{Danqi Chen},
  \bibinfo{person}{Omer Levy}, \bibinfo{person}{Mike Lewis},
  \bibinfo{person}{Luke Zettlemoyer}, {and} \bibinfo{person}{Veselin
  Stoyanov}.} \bibinfo{year}{2019}\natexlab{}.
\newblock \showarticletitle{Roberta: A robustly optimized bert pretraining
  approach}.
\newblock \bibinfo{journal}{\emph{arXiv preprint arXiv:1907.11692}}
  (\bibinfo{year}{2019}).
\newblock


\bibitem[Maldonado-Sifuentes et~al\mbox{.}(2021)]%
        {maldonado2021virality}
\bibfield{author}{\bibinfo{person}{Christian~E Maldonado-Sifuentes},
  \bibinfo{person}{Jason Angel}, \bibinfo{person}{Grigori Sidorov},
  \bibinfo{person}{Olga Kolesnikova}, {and} \bibinfo{person}{Alexander
  Gelbukh}.} \bibinfo{year}{2021}\natexlab{}.
\newblock \showarticletitle{Virality Prediction for News Tweets Using RoBERTa}.
  In \bibinfo{booktitle}{\emph{Mexican International Conference on Artificial
  Intelligence}}. Springer, \bibinfo{pages}{81--95}.
\newblock


\bibitem[Maximilian~Jenders(2013)]%
        {Jenders-2013}
\bibfield{author}{\bibinfo{person}{Felix~Naumann Maximilian~Jenders,
  Gjergji~Kasneci}.} \bibinfo{year}{2013}\natexlab{}.
\newblock \showarticletitle{Analyzing and predicting viral tweets}.
\newblock  (\bibinfo{year}{2013}).
\newblock


\bibitem[Nguyen et~al\mbox{.}(2020)]%
        {nguyen2020bertweet}
\bibfield{author}{\bibinfo{person}{Dat~Quoc Nguyen}, \bibinfo{person}{Thanh
  Vu}, {and} \bibinfo{person}{Anh~Tuan Nguyen}.}
  \bibinfo{year}{2020}\natexlab{}.
\newblock \showarticletitle{BERTweet: A pre-trained language model for English
  Tweets}.
\newblock \bibinfo{journal}{\emph{arXiv preprint arXiv:2005.10200}}
  (\bibinfo{year}{2020}).
\newblock


\bibitem[Rameez et~al\mbox{.}(2022)]%
        {rameez2022viralbert}
\bibfield{author}{\bibinfo{person}{Rikaz Rameez}, \bibinfo{person}{Hossein~A
  Rahmani}, {and} \bibinfo{person}{Emine Yilmaz}.}
  \bibinfo{year}{2022}\natexlab{}.
\newblock \showarticletitle{ViralBERT: A User Focused BERT-Based Approach to
  Virality Prediction}.
\newblock  (\bibinfo{year}{2022}).
\newblock


\bibitem[Samuel et~al\mbox{.}(2020)]%
        {samuel2020message}
\bibfield{author}{\bibinfo{person}{Jim Samuel}, \bibinfo{person}{Myles Garvey},
  {and} \bibinfo{person}{Rajiv Kashyap}.} \bibinfo{year}{2020}\natexlab{}.
\newblock \showarticletitle{That message went viral?! exploratory analytics and
  sentiment analysis into the propagation of tweets}.
\newblock \bibinfo{journal}{\emph{arXiv preprint arXiv:2004.09718}}
  (\bibinfo{year}{2020}).
\newblock


\bibitem[Sprejer et~al\mbox{.}(2021)]%
        {sprejer2021influencer}
\bibfield{author}{\bibinfo{person}{Laila Sprejer}, \bibinfo{person}{Helen
  Margetts}, \bibinfo{person}{Kleber Oliveira}, \bibinfo{person}{David
  O'Sullivan}, {and} \bibinfo{person}{Bertie Vidgen}.}
  \bibinfo{year}{2021}\natexlab{}.
\newblock \showarticletitle{An influencer-based approach to understanding
  radical right viral tweets}.
\newblock \bibinfo{journal}{\emph{arXiv preprint arXiv:2109.07588}}
  (\bibinfo{year}{2021}).
\newblock


\bibitem[Zadeh and Sharda(2022)]%
        {zadeh2022can}
\bibfield{author}{\bibinfo{person}{Amir Zadeh} {and} \bibinfo{person}{Ramesh
  Sharda}.} \bibinfo{year}{2022}\natexlab{}.
\newblock \showarticletitle{How Can Our Tweets Go Viral? Point-Process
  Modelling of Brand Content}.
\newblock \bibinfo{journal}{\emph{Information \& Management}}
  \bibinfo{volume}{59}, \bibinfo{number}{2} (\bibinfo{year}{2022}),
  \bibinfo{pages}{103594}.
\newblock


\end{thebibliography}

\appendix

\end{document}